\documentclass[prd,preprint,nofootinbib]{revtex4-1}

\pdfoutput=1
\usepackage[latin1]{inputenc}
\usepackage[T1]{fontenc}
\usepackage{amsmath}
\usepackage{amssymb}
\usepackage{graphicx}
\usepackage{caption}
\usepackage{subcaption}

\graphicspath{{figures//}}

\usepackage{color}

\begin{document}

\title{Anisotropic shear viscosity of a strongly coupled non-Abelian plasma from magnetic branes}

\author{R.~Critelli}
\email{renato.critelli@usp.br}
\affiliation{Instituto de F\'{i}sica, Universidade de S\~{a}o Paulo, S\~{a}o Paulo, SP, Brazil}

\author{S.~I.~Finazzo}
\email{stefano@if.usp.br}
\affiliation{Instituto de F\'{i}sica, Universidade de S\~{a}o Paulo, S\~{a}o Paulo, SP, Brazil}

\author{M.~Zaniboni}
\email{maiconzs@if.usp.br}
\affiliation{Instituto de F\'{i}sica, Universidade de S\~{a}o Paulo, S\~{a}o Paulo, SP, Brazil}

\author{J.~Noronha}
\email{noronha@if.usp.br}
\affiliation{Instituto de F\'{i}sica, Universidade de S\~{a}o Paulo, S\~{a}o Paulo, SP, Brazil}

\date{\today}

\begin{abstract}

Recent estimates for the electromagnetic fields produced in the early stages of non-central ultra-relativistic heavy ion collisions indicate the presence of magnetic fields $B\sim \mathcal{O}(0.1-15\,m_\pi^2)$, where $m_\pi$ is the pion mass. It is then of special interest to study the effects of strong (Abelian) magnetic fields on the transport coefficients of strongly coupled non-Abelian plasmas, such as the quark-gluon plasma formed in heavy ion collisions. In this work we study the anisotropy in the shear viscosity induced by an external magnetic field in a strongly coupled $\mathcal{N} = 4$ SYM plasma. Due to the spatial anisotropy created by the magnetic field, the most general viscosity tensor of a magnetized plasma has 5 shear viscosity coefficients and 2 bulk viscosities. We use the holographic correspondence to evaluate two of the shear viscosities, $\eta_{\perp} \equiv \eta_{xyxy}$ (perpendicular to the magnetic field) and $\eta_{\parallel} \equiv \eta_{xzxz}=\eta_{yzyz}$ (parallel to the field). When $B\neq 0$ the shear viscosity perpendicular to the field saturates the viscosity bound $\eta_{\perp}/s = 1/(4\pi)$ while in the direction parallel to the field the bound is violated since $\eta_{\parallel}/s < 1/(4\pi)$. However, the violation of the bound in the case of strongly coupled SYM is minimal even for the largest value of $B$ that can be reached in heavy ion collisions. 

\end{abstract}

\maketitle

\section{Introduction}
\label{sec:intro}

The new state of matter formed in ultra-relativistic heavy ion collisions \cite{Adams:2005dq,Adcox:2004mh,Arsene:2004fa,Back:2004je} behaves as a type of strongly coupled Quark-Gluon Plasma (QGP) \cite{Gyulassy:2004zy}. Perhaps one of its most striking features is its apparent near ``perfect'' fluid behavior inferred from comparisons of relativistic hydrodynamic calculations to heavy ion data (for a recent review see \cite{Heinz:2013th}). In fact, the experimental data can be reasonably described \footnote{There are other effects, not included in the analysis of \cite{Heinz:2013th}, which can affect the effective value of $\eta/s$ in the QGP. For instance, there are many transport coefficients in viscous relativistic hydrodynamics \cite{Denicol:2012cn} and very little is known about their values and their effects on the anisotropic flow. In fact, it has been recently found that the inclusion of bulk viscosity directly affects estimates of $\eta/s$ in the QGP \cite{Noronha-Hostler:2013gga,Noronha-Hostler:2014dqa}.} using very small values of the shear viscosity to entropy density ratio, $\eta/s\sim 0.2$ \cite{Heinz:2013th}, which is of the order of the ratio $\eta/s=1/(4\pi)$ \cite{Policastro:2001yc,Buchel:2003tz,Kovtun:2004de} found in a large class of strongly coupled non-Abelian plasmas using the gauge/gravity duality \cite{Maldacena:1997re,Gubser:1998bc,Witten:1998qj} (see \cite{CasalderreySolana:2011us} for a review that includes applications to heavy ion collisions). Such a small $\eta/s$ is not really compatible with standard weak coupling QCD results \cite{Arnold:2000dr,Arnold:2003zc} and other mechanisms/models have been tried over the years to explain this ratio \cite{Danielewicz:1984ww,Asakawa:2006tc,Meyer:2007ic,Xu:2007ns,Hidaka:2008dr,NoronhaHostler:2008ju,NoronhaHostler:2012ug,Asakawa:2012yv,Ozvenchuk:2012kh}. In this aspect, the gauge/gravity duality remains as one of the leading non-perturbative tools suited for calculations of real time properties of strongly coupled non-Abelian plasmas.  

In the last few years, several works have emphasized that non-central heavy ion collisions are not only characterized by a sizable anisotropic flow but also by the presence of very strong electromagnetic fields formed at the early stages of the collisions \cite{Kharzeev:2007jp,Fukushima:2008xe,Skokov:2009qp,Tuchin:2013ie,Deng:2012pc,Bloczynski:2012en}. This has sparked a lot of interest on the effects of strong electromagnetic fields in strongly interacting QCD matter \cite{bookdima} and, recently, lattice calculations with physical quark masses have determined how a strong external magnetic field changes the thermodynamic properties of the QGP \cite{Bali:2011qj,Bali:2014kia}. Lattice calculations have also been used in \cite{Bonati:2013lca,Bonati:2013vba,Bali:2013owa} to determine the magnetization of QCD matter in equilibrium and the authors of Ref.\ \cite{Bali:2013owa} argued that the paramagnetic behavior \cite{Fraga:2012ev} found in these lattice simulations leads to a sort of paramagnetic squeezing that could contribute to the overall elliptic flow observed in heavy ion collisions. If the magnetic field is still large enough at the time that elliptic flow is building up, it is natural to also consider the effects of strong magnetic fields on the subsequent hydrodynamic expansion of the QGP. 

\hspace{1cm}

The strong magnetic field breaks the spatial $SO(3)$ rotational symmetry to a $SO(2)$ invariance about the magnetic field axis and this type of magnetic field-induced anisotropic relativistic hydrodynamics has more transport coefficients than the more symmetric case in order to distinguish the dynamics along the magnetic field direction from that in the plane orthogonal to the field. In fact, this means that the number of independent transport coefficients in the shear viscosity tensor $\eta_{ijkl}$ increases from 1 (in the isotropic case) to 5 in the presence of the magnetic field while there are 2 bulk viscosity coefficients \cite{Huang:2011dc,LandauKine,LandauEla,Tuchin:2011jw}. Therefore, one needs to know how this ``Zeeman-like'' splitting of the different viscosity coefficients depends on the external magnetic field to correctly assess the phenomenological consequences of strong fields on the hydrodynamic response of the QGP formed in heavy ion collisions.

Since one no longer has $SO(3)$ invariance, one may expect that some of the different shear viscosities could violate the universal result $\eta/s = 1/(4\pi)$ valid for isotropic Einstein geometries \cite{Buchel:2003tz,Kovtun:2004de}, which would then constitute an example of the violation of the viscosity bound that is of direct relevance to heavy ion collisions. Previous examples involving the violation of the viscosity bound include: anisotropic deformations of $\mathcal{N} = 4$ Super-Yang-Mills (SYM) theory due to a $z$-dependent axion profile \cite{Mateos:2011ix}  computed in \cite{Rebhan:2011vd} where $\eta_{\parallel}/s< 1/(4\pi)$ along the direction of anisotropy; anisotropic holographic superfluids with bulk $SU(2)$ non-Abelian fields which present universality deviation for $\eta_{\parallel}/s$ \cite{Natsuume:2010ky,Erdmenger:2010xm,Erdmenger:2012zu}; and a dilaton-driven anisotropic calculation recently shown in \cite{Jain:2014vka}. We remark, however, that the first examples of viscosity bound violation were found in ($SO(3)$ invariant) theories with higher order derivatives in the gravity dual \cite{Kats:2007mq,Brigante:2007nu,Brigante:2008gz,Buchel:2008vz}.

In this paper we evaluate two components of the shear viscosity tensor, namely $\eta_{\perp} \equiv \eta_{xyxy}$ and $\eta_{\parallel} \equiv \eta_{xzxz}=\eta_{yzyz}$, in a strongly coupled non-Abelian plasma in the presence of an external magnetic field using the gauge/gravity duality (other two shear coefficients are identically zero for the theory considered here, as shown in the Appendix). These calculations are done using the membrane paradigm \cite{Iqbal:2008by,Thorne:1986iy}. The holographic model we consider is simple Einstein gravity (with negative cosmological constant) coupled with a (prescribed) Maxwell field, which correspond to strongly coupled $\mathcal{N} = 4$ SYM subjected to an external constant and homogenous magnetic field \cite{D'Hoker:2009bc,D'Hoker:2009mm,D'Hoker:2010ij}. We examine the role played by the anisotropy introduced by the external field searching for a violation of the viscosity bound in $\eta_{\parallel}/s$. A study of the behavior of $\eta_{\parallel}/s$ is also of phenomenological interest for the modeling of the strongly coupled QGP under strong magnetic fields.

This work is organized as follows. In Section \ref{sec:magbranes} we review the thermodynamics of the magnetic brane background found in \cite{D'Hoker:2009mm}, introduce our notation, and discuss the numerical procedure used to solve the Einstein-Maxwell coupled equations. In Section \ref{sec:shear}, after a preliminary discussion about the computation of $\eta/s$ from the membrane paradigm in isotropic theories, we show that metric fluctuations in this background parallel and transverse to the external magnetic field result in scalar field fluctuations with two different couplings. This result can then be used in the context of the membrane paradigm to evaluate the shear viscosity coefficients $\eta_\perp$ and $\eta_\parallel$. We finish in Section \ref{sec:conc} with a discussion of our results.

\section{Magnetic brane background}
\label{sec:magbranes}


We consider in the bulk a simple Einstein+Maxwell system and look for solutions corresponding to the deformation of the $AdS_5$-Schwarzchild geometry due to a $\mathrm{U(1)}$ Abelian gauge field \cite{D'Hoker:2009mm}. The $\mathrm{U(1)}$ gauge field is chosen to give a constant and homogeneous magnetic field. This magnetic field in the bulk is then taken as an external magnetic field at the boundary gauge theory \cite{D'Hoker:2009mm}, which is strongly coupled $\mathcal{N}=4$ SYM. Clearly, the adjoint fermions in SYM feel directly the effects of the magnetic field but, due to fermion loops, the gluon sector is also affected by the field. This is why the thermodynamic properties of this ``magnetic'' SYM plasma considerably differ from those found in SYM in the absence of external fields. 

Let us review this background and its thermodynamic properties. The action of the 5-dimensional gravitational bulk theory is given by the Einstein-Hilbert action coupled with a Maxwell field
\begin{equation}
\label{eq:action}
S = \frac{1}{16 \pi G_5} \int d^5x \, \sqrt{-g}  \left(R + \frac{12}{L^2} - F^2\right) + S_{CS} + S_{GH},
\end{equation}
where $G_5$ is the 5-dimensional gravitational constant, $L$ is the asymptotic $AdS_5$ radius and $F$ is the Maxwell field strength 2-form. The terms $S_{CS}$ and $S_{GH}$ are the Chern-Simons and Gibbons-Hawking terms. The latter is necessary to define a well posed variational problem but both the CS and GH terms will not play a role in the calculation of shear viscosity coefficients\footnote{We note that our definition for the Riemann tensor possesses an overall minus sign in comparison to the one used in \cite{D'Hoker:2009mm}.}. Other terms are needed in (\ref{eq:action}) from the viewpoint of holographic renormalization but those do not affect the calculations performed in this paper. The equations of motion are then given by the Einstein's equations
\begin{equation}
\label{eq:eom1}
R_{\mu \nu} = -\frac{4}{L^2} g_{\mu \nu} - \frac{1}{3} F_{\rho \sigma} F^{\rho \sigma} g_{\mu \nu} + 2 F_{\mu \rho} F_{\nu}^{\ \rho},
\end{equation}
and the Maxwell's field equations for the Abelian field,
\begin{equation}
\label{eq:eom2}
\nabla_{\mu} F^{\mu \nu} = 0.
\end{equation}
Following \cite{D'Hoker:2009mm}, the Ansatz for the magnetic brane geometry is 
\begin{equation}
\label{eq:background}
ds^2 = -U(r) dt^2 + \frac{dr^2}{U(r)} + f(r)(dx^2+dy^2) + p(r) dz^2,
\end{equation}
where $U(r)$, $f(r)$ and $p(r)$ are determined by solving the equations of motion. The holographic coordinate $r$ is such that the boundary is located at $r \to \infty$. We want a black brane background and, thus, we require that at a given $r=r_h$ the function $U(r)$ has a simple zero. The Ansatz for the field strength $F$ is given by
\begin{equation}
\label{eq:fieldstrength}
F = B\, dx \wedge dy,
\end{equation}
where the constant $B$ is the bulk magnetic field, oriented along the $z$ direction. It can be checked that the equation of motion \eqref{eq:eom2} is trivially satisfied by this Ansatz.

In the absence of a magnetic field $p(r) = f(r)$, which reflects the spatial $SO(3)$ invariance of the boundary gauge theory. However, since the magnetic field establishes a preferred direction in space, it breaks the $SO(3)$ spatial symmetry to only a $SO(2)$ symmetry in the $x,y$ directions. In the bulk theory this is taken into account by the fact that in this case $f(r) \neq p(r)$.

The equations of motion derived from \eqref{eq:background} are
\begin{align}
\label{eq:eom3}
 U(V''-W'') + \left(U'+U(2V'+W') \right)(V'-W') & =  -2B^2 e^{-4V}, \nonumber \\
2 V'' + W'' + 2 (V')^2 + (W')^2 & =  0, \nonumber \\
\frac{1}{2} U'' + \frac{1}{2} U' (2V'+W') & =  4 + \frac{2}{3} B^2 e^{-4V} \quad \quad \\
2 U' V' + U'V + 2U(V')^2+4UV'W' & =  12 - 2B^2 e^{-4V}, \nonumber
\end{align}
where we defined $V$ and $W$ by $f = e^{2V}$ and $p = e^{2W}$. By Bianchi's identity, the fourth equation of motion can be shown to be a consequence of the three first equations and, thus, it can be taken as a constraint on initial data.

It is well-known that charged systems undergo dimensional reduction in the presence of strong fields due to the projection towards the lowest Landau level \cite{Gusynin:1994re,Gusynin:1994xp,Gusynin:1995nb} (see the recent review in \cite{Shovkovy:2012zn}). Taking that into account, the authors of \cite{D'Hoker:2009mm} proposed that the background \eqref{eq:background} satisfied two conditions. The first condition is that the geometry must be asymptotically $\mathrm{AdS}_5$, that is, $U(r) \to r^2$, $p(r) \to r^2$ and $f(r) \to r^2$ when $r \to \infty$ since in the UV we must recover the dynamics of $\mathcal{N} = 4$ SYM without the influence of the magnetic field. The second condition is that in the asymptotic IR the geometry becomes a BTZ black hole \cite{Banados:1992wn} times a two dimensional torus $T^2$ in the spatial directions orthogonal to the magnetic field. In fact, deep in the IR the geometry near the horizon of the black brane $r_h$, $r \sim r_h$, is given by
\begin{equation}
\label{eq:BTZ}
ds^2 = \left[ -3(r^2-r_h^2) dt^2 + 3 r^2 dz^2 + \frac{dr^2}{3(r^2-r_h^2)} \right] + \left[ \frac{B}{\sqrt{3}} (dx^2+dy^2)\right].
\end{equation}
This implies that in the IR the dynamics corresponds to a (1+1) dimensional CFT. Thus, imposing that the background interpolates between the BTZ black hole for $r \sim r_h$ and $\mathrm{AdS}_5$ for high $T$ and interpreting the flow along the $r$ direction as a renormalization group flow, this solution flows from a (1+1) dimensional CFT in the IR to a 4 dimensional CFT in the UV \cite{D'Hoker:2009mm}.

\subsection{Numerical solution and thermodynamics}
\label{sec:num}

Unfortunately, no analytic solution which interpolates between $AdS_5$ and the BTZ$\times T^2$ geometry is known and, thus, we must resort to numerics. In this subsection we briefly review the numerical procedure for solving the equations of motion and the thermodynamics, first elaborated in \cite{D'Hoker:2009mm}. We do so since the same procedure will be used to determine $\eta_{//}/\eta_{\perp}$ numerically in Section \ref{sec:shear}. 

The strategy is to first choose the scale for the $t$ and $r$ coordinates to fix the horizon position at $r_h=1$ so that $\tilde{U}(1) = 0$, where the tilde indicates that we are in the rescaled coordinates $\tilde{t}$ and $\tilde{r}$. By using the fact that any physical quantity in this model should depend on the dimensionless ratio $T/\sqrt{B}$, we also fix the temperature at $T=1/(4 \pi)$ - this means that we take $\tilde{U}'(1) = 1$. Also, we rescale the $x$, $y$, and $z$ coordinates to have $\tilde{V}(1)=\tilde{W}(1)=0$. In these new coordinates, the magnetic field is $b$. After these redefinitions, the first and fourth equations in \eqref{eq:eom3} imply that 
\begin{align}
\label{eq:initialdata}
\tilde{V}'(1) = & \,4-\frac{4}{3} b^2 \quad \quad \mathrm{and} \nonumber \\ 
\tilde{W}'(1) = & \, 4+\frac{2}{3}b^2.
\end{align}

This gives a well posed initial value problem for $\tilde{U}(\tilde{r})$, $\tilde{V}(\tilde{r})$, and $\tilde{W}(\tilde{r})$, which can be integrated out from $\tilde{r}=1$ to a large value of $\tilde{r}$. It can be checked numerically that the geometry has the asymptotic behavior
\begin{equation}
\tilde{U}(\tilde{r})\rightarrow \tilde{r}^2, \ \ \ e^{2\tilde{V}(\tilde{r})} \rightarrow v \tilde{r}^2, \ \ \ e^{2\tilde{W}(\tilde{r})} \rightarrow w \tilde{r}^2,
\end{equation}
where $v(b)$ and $w(b)$ are proportionality constants that depend on the rescaled magnetic field $b$. This result implies that, apart from a coordinate rescaling, the geometry is asymptotically $\mathrm{AdS}_5$.  To go back to the original units and have the correct $\mathrm{AdS}_5$ asymptotic behavior, we need to rescale back to our original coordinate system by doing $(\tilde{x},\tilde{y},\tilde{z})\rightarrow(x/\sqrt{v},y/\sqrt{v},z/\sqrt{w})$. The metric is then (in coordinates that are asymptotically $\mathrm{AdS}_5$)
\begin{equation}
\label{eq:resc.metric}
ds^2 = -\tilde{U}(r)dt^2 + \frac{dr^2}{\tilde{U}(r)} + \frac{e^{2\tilde{V}(r)}}{v}(dx^2+dy^2) + \frac{e^{2\tilde{W}(r)}}{w}dz^2,
\end{equation}
where we note that we have taken $r = \tilde{r}$. By the same token, the field strength is now written as
\begin{equation}
F = \frac{b}{v}dx \wedge dy.
\end{equation}
Therefore, the rescaled magnetic field is related to the physical field at the boundary by $B = b/v$. Also, note that the first equation \eqref{eq:initialdata} implies that for $b > \sqrt{3}$ we have $V'(1) < 0$, which means that the geometry will not be asymptotically $\mathrm{AdS}_5$. Thus, the rescaled field $b$ has an upper value given by $b_{max} = \sqrt{3}$.

From \eqref{eq:resc.metric}, one can obtain the thermodynamics of the gauge theory. The physical field is $\mathcal{B} = \sqrt{3} B$, as argued in \cite{D'Hoker:2009mm} by comparing the Chern-Simons term in \eqref{eq:action} with the $\mathcal{N} = 4$ SYM chiral anomaly. The dimensionless ratio $T/\sqrt{\mathcal{B}}$ is given by
\begin{equation}
\frac{T}{\sqrt{\mathcal{B}}} = \frac{1}{4 \pi\, 3^{1/4}} \sqrt{\frac{v}{b}}.
\end{equation}
while the dimensionless ratio of the entropy density $s$ by $N^2 \mathcal{B}^3/2$ (using that $G_5 = \pi/2N^2$) is
\begin{equation}
\frac{s}{N^2\mathcal{B}^{3/2}} = \frac{1}{3^{3/4} 2\pi}\sqrt{\frac{v}{b^3 w}}\,.
\end{equation}

\begin{figure}
\centering
  \includegraphics[width=.6\linewidth]{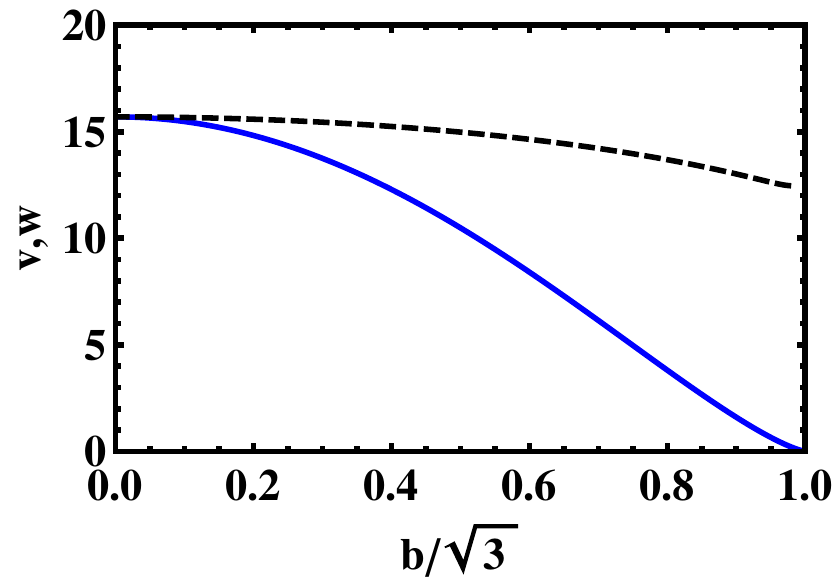}
  \caption{(Color online) The rescaling parameters $v$ (solid blue curve) and $w$ (dashed black curve) as a function of $b/\sqrt{3}$.}
  \label{fig:vwb}      
\end{figure}

\begin{figure}
\centering
  \includegraphics[width=.6\linewidth]{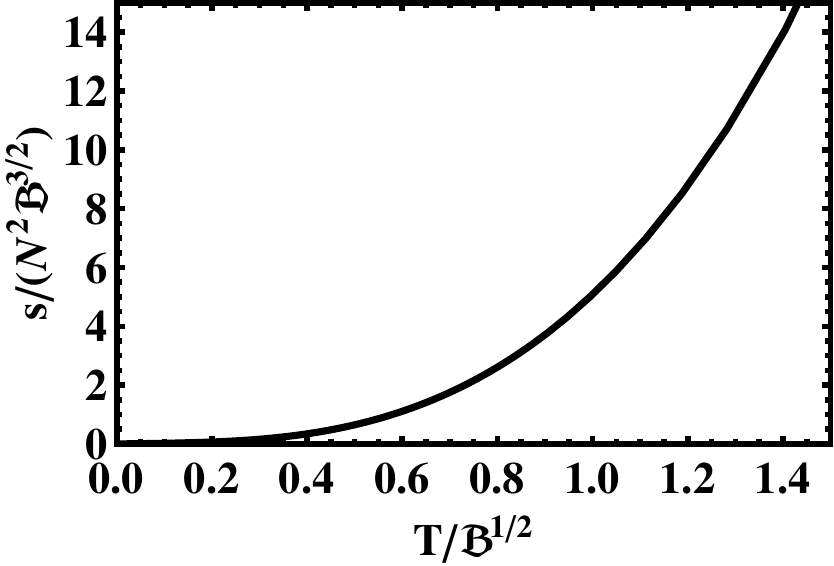}
  \caption{The normalized entropy density $s/(N^2 \mathcal{B}^{3/2})$ as a function of the dimensionless combination $T/\sqrt{\mathcal{B}}$.}
  \label{fig:thermo}      
\end{figure}

The numerical procedure for evaluating the thermodynamics can then be summarized as follows: one chooses a value of the rescaled magnetic field $b$, numerically solves the equations of motion, and obtains the rescaled parameters $v$ and $w$ by fitting the asymptotic data for $\tilde{V}(r)$ and $\tilde{W}(r)$ to the functions $v r^2$ and $w r^2$. By varying $b$, one can obtain the functions $v(b)$ and $w(b)$ and evaluate $T/\sqrt{\mathcal{B}}$ versus $s/(N^2 \mathcal{B}^{3/2})$ by using $b$ as a parameter. In Fig.\ \ref{fig:vwb} we show $v$ and $w$ as a function of $b$. The entropy density is shown in Fig.\ \ref{fig:thermo} and we have checked that our results match those previously found in \cite{D'Hoker:2009mm}.

\section{Anisotropic shear viscosity due to an external magnetic field}
\label{sec:shear}

\subsection{Isotropic shear viscosity}

From linear response theory \cite{Kapusta:2006pm}, the viscosity tensor for an anisotropic theory is given by the Kubo formula
\begin{equation}
\label{eq:sheartensor}
\eta_{ijkl} = - \lim_{\omega \to 0} \frac{1}{\omega} \text{Im} \ G_{ij,kl}^{R} (\omega, \vec{k} = 0) \ \ \text{with} \ i, j,k, l = x, y, z
\end{equation}
where $G_{ij,kl}^{R}(\omega, \vec{k})$ is the momentum space retarded Green's function given by
\begin{equation}
G_{ij,kl}^{R}(\omega, \vec{k}) = -i \int d^4x \, e^{-i k \cdot x} \theta(t)\left\langle \left[ \hat{T}_{ij}(x), \hat{T}_{kl}(0) \right] \right\rangle,
\end{equation}
while $\hat{T}_{ij}$ is the stress energy operator in the quantum field theory.

For an isotropic theory of hydrodynamics in the absence of other conserved currents, there are only two transport coefficients associated with energy and momentum at the level of relativistic Navier-Stokes theory, namely the isotropic shear viscosity $\eta$ and the bulk viscosity $\zeta$. The computation of $\eta$ in strongly coupled gauge theories using the gauge/gravity duality, in the case of isotropic gauge theories with two derivative gravitational duals, gives a universal value \cite{Policastro:2001yc,Kovtun:2004de}
\begin{equation}
\label{eq:etasiso}
\frac{\eta}{s} = \frac{1}{4\pi}.
\end{equation}

A convenient method that can be used to derive this result is the membrane paradigm \cite{Iqbal:2008by}. In this framework, if we want to compute the transport coefficient $\chi$ of a scalar operator $\hat{O}$ given by the Kubo formula
\begin{equation}
\chi = - \lim_{\omega \to 0} \frac{1}{\omega} \text{Im} \ G^{R} (\omega, \vec{k} = 0),
\end{equation}
where $G^R$ is the retarted correlator associated with the scalar operator $\hat{O}$
\begin{equation}
G^{R}(\omega, \vec{k}) = -i \int d^4x \, e^{-i k \cdot x} \theta(t)\langle \left[ \hat{O} (x), \hat{O} (0) \right] \rangle,
\end{equation}
one needs to look for fluctuations $\phi$ of the associated bulk field in dual gravity theory, in accordance with the gauge/gravity dictionary \cite{Gubser:1998bc,Son:2002sd}. In the case that the action for the fluctuations is given by a massless scalar field with an $r$ dependent coupling $\mathcal{Z}(r)$,
\begin{equation}
S_{fluc} = - \int d^5x \, \sqrt{-g} \frac{1}{2 \mathcal{Z}(r)} (\partial \phi)^2 ,
\end{equation}
the transport coefficient $\chi$ is given by the corresponding transport coefficient $\chi_{mb}$ of the stretched membrane of the black brane horizon \cite{Iqbal:2008by}
\begin{equation}
\label{eq:chi}
\chi = \chi_{mb} = \frac{1}{\mathcal{Z}(r_h)}.
\end{equation}
In the case of the isotropic shear viscosity $\eta$, we must consider the fluctuations $h_{xy}$ of the metric component $g_{xy}$ since the energy-momentum tensor operator in the gauge theory $\hat{T}_{\mu \nu}$ is dual to the bulk metric $g_{\mu \nu}$ of the gravity dual. Given that in isotropic backgrounds the mixed fluctuation $h_{x}^y$ can be described as the fluctuation of a massless scalar field with $\mathcal{Z}(r) = 16 \pi G_5$ \cite{Kovtun:2004de}, then $\eta = 1/(16 \pi G_5)$. The universal result in \eqref{eq:etasiso} follows from identifying the entropy density with the area of the horizon via the Bekenstein formula.

\subsection{Metric fluctuations and anisotropic shear viscosity}

Let us now consider metric fluctuations about the background \eqref{eq:background}, which is a solution of the Einstein-Maxwell system \eqref{eq:action}. In a fluid with axial symmetry about an axis due to an external magnetic field there are, in principle, 7 independent transport coefficients in the full viscosity tensor $\eta_{ijkl}$ defined in \eqref{eq:sheartensor}, five of which are shear viscosities and the other two bulk viscosities \cite{LandauKine,Huang:2011dc} - for completeness, in the Appendix we present a brief derivation of this result. However, as also shown in the Appendix, of the five shear viscosities, two of them are identically zero for the class of anisotropic diagonal backgrounds given by Eq. \eqref{eq:background}, which reduces the total number of independent components of the shear tensor from 7 to 5 (incidentally, anisotropic superfluids also have 5 transport coefficients \cite{LandauEla,Erdmenger:2010xm}). In our case, we are especially interested in the following two components of $\eta_{ijkl}$,
\begin{equation}
 \eta_{xyxy} = \eta_{\perp}, \ \ \ \text{and} \ \ \ \eta_{yzyz}= \eta_{xzxz} =\eta_{\parallel}\,.
\end{equation}

The magnetic field breaks the $SO(3)$ rotational invariance of background to only a $SO(2)$ rotation invariance about the $z$ axis. Thus, as expected, it is possible to show that linearized $\phi(t,r) = h_x^y(t,r)$ fluctuations obey
\begin{equation}
\label{eq:gravityflucfin}
\delta S = -\frac{1}{32 \pi G_5} \int d^5x \sqrt{-g}\, (\partial \phi)^2,
\end{equation}
which means that the shear viscosity $\eta_{xyxy} \equiv \eta_{\perp}$ is still given by \eqref{eq:etasiso} and this shear coefficient saturates the viscosity bound.

However, $h_{zx}$ (or, equivalently, $h_{zy}$) fluctuations are not protected by the remaining rotation invariance of the background. In fact, in the context of the membrane paradigm, we must first show that the fluctuation $h_{zx}(t,r)$ obeys the equation of a massless scalar field in order to apply \eqref{eq:chi}. However, the coupling in the action may differ from \eqref{eq:gravityflucfin} and, thus, $\eta_{\parallel} \neq \eta_{\perp}$.

Consider then a fluctuation of the form $g_{zx} \to g_{zx} + h_{zx}$ \footnote{One can show that homogeneous fluctuations of the $\mathrm{U(1)}$ bulk field $A_{\mu}$ decouple from the corresponding fluctuations $h_{xy}$ and $h_{zx}$.}. In order to have a scalar-like action with just the kinetic term (and possibly an $r$ dependent coupling), we choose the mode $\psi(t,r) \equiv h_{y}^{z}(t,r) $, rather than $h_z^y $ for example. Inserting this fluctuation into the action and keeping only quadratic terms one can show that
\begin{align}
\delta S = & \frac{1}{16 \pi G_5}  \int \sqrt{-g} \left\{ \psi^2 \left[ \frac{\square p}{f} -\frac{p}{f^2}\square f -\frac{3}{2f^2}\partial_\mu f\partial^\mu p +\frac{3p}{2f^3}(\partial f)^2  \right]  \right. + \nonumber \\
+ & \left[ \frac{2p}{f}\psi\square\psi - \frac{3p}{2f^2}\partial_\mu f\partial^\mu\psi^2+\frac{2}{f}\partial_\mu p\partial^\mu\psi^2 \right] + \nonumber \\
+ & \left[-\frac{3p}{2f}\frac{(\partial_t\psi)^2}{U} + \frac{3p}{2f}U(\partial_r\psi)^2 = \frac{3p}{2f}\partial_\mu\psi\partial^\mu\psi \right] + \\
- & \left.\left[\left(R+\frac{12}{L^2} -F^2 \right)\frac{p}{2f}\psi^2 +\frac{p}{f}F^2\psi^2 \right] \right\}, \nonumber
\end{align}
where the d'Alembertian is 
\begin{equation}
\Box = -\frac{1}{U}\partial_t^2 + U\partial_r^2+\left(U'+\frac{Uf'}{f}+\frac{Up'}{2p}  \right)\partial_r\,.
\end{equation}
Now, using that the trace of the Einstein's equations gives $R + 20/L^2 = F^2/3$ and, integrating by parts the $\psi \Box \psi$ term, we obtain
\begin{align}
\label{eq:maxfluc}
\delta S &  = \frac{1}{16 \pi G_5}  \int d^5x \, \sqrt{-g} \left[- \frac{p}{2f}\partial_\mu\psi\partial^\mu\psi  - \frac{p}{2f^2}\partial_\mu f\partial^\mu\psi^2+\frac{1}{f}\partial_\mu p\partial^\mu\psi^2 + \right. \nonumber \\ & \left. + \psi^2 \left( \frac{\square p}{f} -\frac{p}{f^2}\square f -\frac{3}{2f^2}\partial_\mu f\partial^\mu p +\frac{3p}{2f^3}(\partial f)^2 \right) + \left(\frac{4p}{fL^2}\psi^2+\frac{F^2}{3}\frac{p}{f}\psi^2  \right) -\frac{p}{f}F^2\psi^2  \right].
\end{align}
We now use the unperturbed Einstein's equations. One needs the $zz$ equation
\begin{equation}
\label{eq:einsteinzz}
\frac{4p}{fL^2} = \frac{\square p}{2f} - \frac{(\partial p)^2}{2pf} - \frac{F^2}{3}\frac{p}{f}
\end{equation}
and also the $yy$ equation,
\begin{equation}
\label{eq:einsteinyy}
-\frac{1}{2}\square p +\frac{(\partial p)^2}{2p} = -\frac{4p}{L^2}-\frac{F^2}{3}p.
\end{equation}
Using the $zz$ \eqref{eq:einsteinzz} equation in \eqref{eq:maxfluc} and integrating by parts once again, noting that
\begin{equation}
\frac{1}{f}\partial_\mu p \partial^\mu\psi^2 = \nabla_\mu\left( \frac{\partial^\mu p}{f}\psi^2 \right)+\frac{1}{f^2}\partial_\mu f\partial^\mu p \psi^2 -\psi^2\frac{\square p}{f} \quad \quad \mathrm{and}
\end{equation}
\begin{equation}
-\frac{p}{2f^2}\partial_\mu f\partial^\mu\psi^2 = - \nabla_\mu\left(\psi^2\frac{p}{2f}\partial^\mu f \right) +\frac{\psi^2}{2f^2}\partial_\mu p\partial^\mu f-\frac{p}{f^3}\psi^2(\partial f)^2 +\frac{p}{2f^2}\psi^2 \square f,
\end{equation}
we arrive at
\begin{align}
\label{eq:maxfluc2}
\delta S = & \frac{1}{16\pi G_5}  \int d^5x \, \sqrt{-g}\left[- \frac{p}{2f}\partial_\mu\psi\partial^\mu\psi +\frac{p}{f}\psi^2 \right. \nonumber + \\ & + \left. \frac{p}{f} \psi^2 \left( \frac{1}{2}\frac{\square p}{p} -\frac{1}{2f}\square f  +\frac{1}{2f^2}(\partial f)^2 - \frac{(\partial p)^2}{2p^2} \right) -\frac{p}{f}F^2\psi^2 \right]\,.
\end{align}
Finally, from \eqref{eq:einsteinzz} and \eqref{eq:einsteinyy}
\begin{equation}
\frac{1}{2}\frac{\square p}{p} -\frac{1}{2f}\square f  +\frac{1}{2f^2}(\partial f)^2 - \frac{(\partial p)^2}{2p^2} = F^2,
\end{equation}
one can show that the action for the fluctuations \eqref{eq:maxfluc2} becomes
\begin{equation}
\label{eq:maxfluc3}
\delta S =  - \frac{1}{16\pi G_5}  \int d^5x \, \sqrt{-g}\left(\frac{p(r)}{2f(r)}\partial_\mu\psi\partial^\mu\psi \right).
\end{equation}
Therefore, we have a massless scalar field with an $r$ dependent coupling $\mathcal{Z}(r) = 16\pi G_5f(r)/p(r)$. These functions were found in the previous section to determine the thermodynamic properties of this system and, thus, in the next section we shall evaluate $\eta_\parallel$. 

\subsection{Viscosity bound violation due to an external magnetic field}

\begin{figure}
\centering
  \includegraphics[width=.65\linewidth]{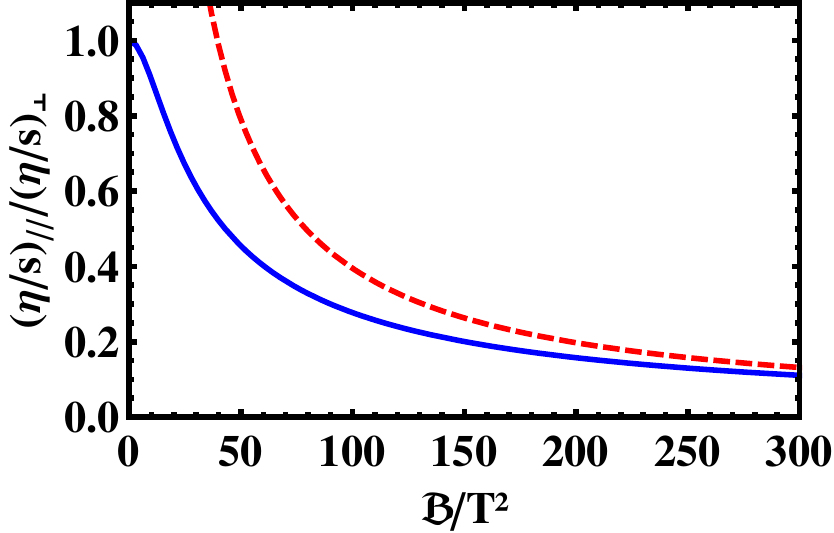}
  \caption{(Color online) The ratio of shear viscosities $(\eta/s)_{\parallel}/(\eta/s)_{\perp}$ as a function of $\mathcal{B}/T^2$. The solid blue line is the numerical result from $(\eta/s)_{\parallel}/(\eta/s)_{\perp} = w/v$; the dashed red line is the asymptotic result valid only when $\mathcal{B} \gg T^2$. \eqref{eq:high}}
  \label{fig:etaaniso}      
\end{figure}

From the result of the previous section, it follows that we can also apply the membrane paradigm to \eqref{eq:maxfluc3} to evaluate $\eta_{\parallel}$, using \eqref{eq:chi}. We then have
\begin{equation}
\frac{\eta_{\parallel}}{s} = \frac{1}{4\pi} \frac{p(r_h)}{f(r_h)}.
\end{equation}
In terms of the numerical, rescaled geometry described in \eqref{eq:resc.metric}, we then obtain
\begin{equation}
\frac{\eta_{\parallel}}{s} = \frac{1}{4\pi} \frac{w}{v}.
\end{equation}
Thus, the ratio $(\eta/s)_{\parallel}/(\eta/s)_{\perp}$ is given by $w/v$. Using this result, we can then evaluate the degree of anisotropy of the shear viscosities as a function of $\mathcal{B}/T^2$; we show the results in Fig.\ \ref{fig:etaaniso}. One can see that for $\mathcal{B}/T^2 \ll 1$, $\eta_{\parallel} \to \eta_{\perp}$, reflecting the fact that at high temperatures we recover the isotropic strongly coupled SYM plasma limit. The asymptotic behavior in the opposite limit, $\mathcal{B}/T^2 \gg 1$, can be understood by looking at the BTZ metric \eqref{eq:BTZ}, which is the relevant geometry in this case. Evaluating $\eta_{\parallel}$ in this limit, one obtains the asymptotic behavior
\begin{equation}
\label{eq:high}
\frac{\eta_{\parallel}}{s} \sim \pi \frac{T^2}{\mathcal{B}}, \ \ \ (\mathcal{B} \gg T^2),
\end{equation}
which is also shown in Fig.\ \ref{fig:etaaniso}. We should note that in this model, $\eta_{\parallel}/s < 1/(4\pi)$ whenever $B>0$. This gives another example in which the viscosity bound in a gravity dual is violated due to anisotropy. The formula above indicates that $\eta_{\parallel}/s$ can become much smaller than $1/(4\pi)$ for sufficiently strong fields. However, it is conceivable that in this limit other constraints must be imposed to obtain a well defined theory. In fact, it was found in \cite{Brigante:2007nu,Brigante:2008gz} that causality in the gauge theory constituted an important constraint that was used to set a lower value for $\eta/s$ in that particular case involving higher order derivatives in the gravity dual. This matter deserves further study and we hope to address this question in the future. 

\section{Conclusions}
\label{sec:conc}

Motivated by the recent studies involving the effects of electromagnetic fields on the strongly coupled plasma formed in heavy ion collisions, in this paper we used the holographic correspondence to compute two anisotropic shear viscosity coefficients of a strongly coupled $\mathcal{N}=4$ SYM plasma in the presence of a magnetic field. As expected, the shear viscosity that describes the dynamics in the plane transverse to the magnetic field, $\eta_\perp$ is not affected by the field and, thus, it still saturates the viscosity bound, i.e., $\eta_\perp/s=1/(4\pi)$. On the other hand, the shear viscosity coefficient along the axis parallel to the external field, $\eta_\parallel$, violates the bound when $B>0$. In fact, we find $\eta_\parallel/s < 1/(4\pi)$. These results are qualitatively similar to those found in \cite{Rebhan:2011vd} for the case of an anisotropic plasma created by a spatial dependent axion profile \cite{Mateos:2011ix}. However, the source of anisotropy in our case (the magnetic field) is arguably more directly connected to heavy ion phenomenology than the one used in \cite{Rebhan:2011vd}. 

Plasmas in the presence of magnetic fields usually experience instabilities and it would be interesting to investigate whether there are instabilities induced by strong magnetic fields in the strongly coupled plasma studied in this paper. In fact, one could compute the spectral functions and the quasi-normal modes associated with $\eta_\parallel$ and check if there is any sudden change in their behavior at strong fields. Also, instabilities in homogenous magnetic media can sometimes be resolved by the formation of magnetic domains and, thus, it would be interesting to investigate whether this is the case for the theory considered in this paper.

Our results for the magnetic field dependence of $\eta_\parallel/s$ show that this ratio only deviates significantly from $1/(4\pi)$ when $\mathcal{B}/ T^2 \gg 1$. Taking the typical temperature at the early stages of heavy ion collisions to be $T \sim 2 m_\pi$, we see that $4\pi\eta_\parallel/s \sim 0.9$ when $\mathcal{B} \sim 40 m_\pi^2$. This value of magnetic field may be too large for heavy ion phenomenology and, thus, our results suggest that anisotropic shear viscosity effects in strongly coupled plasmas are minimal and the isotropic approximation is justified. It would be interesting to check if the same behavior is obtained in strongly coupled plasmas that are not conformal (such as the bottom-up models in Refs.\ \cite{Gursoy:2007cb,Gursoy:2007er,Gursoy:2008bu,Gubser:2008ny,Noronha:2009ud}) to see if there is some nontrivial interplay between the confinement/deconfinement scale and the external magnetic field. Such a study would perhaps give a better idea of the magnetic field induced-anisotropy in the shear viscosity of the QGP. Alternatively, one could also study the effects of strong magnetic fields on the weak coupling calculations of \cite{Arnold:2000dr,Arnold:2003zc} perhaps following the general procedure to compute transport coefficients of relativistic hydrodynamics from the Boltzmann equation proposed in \cite{Denicol:2011fa}.

\acknowledgments

This work was supported by Funda\c c\~ao de Amparo \`a Pesquisa do Estado de
S\~ao Paulo (FAPESP),  Coordena\c c\~ao de Aperfei\c coamento de Pessoal de N\'ivel Superior (CAPES), and Conselho Nacional de Desenvolvimento Cient\'ifico e Tecnol\'ogico (CNPq). The authors thank R.~Rougemont for fruitful discussions on metric fluctuations in asymptotic AdS spacetimes.

\appendix

\section{Shear tensor in a magnetic field}

In this Appendix we show how one can determine the form of the shear tensor in the presence of an external magnetic field - the detailed discussion can be found in \cite{Huang:2011dc}. Here we will present an overview of how one can construct a rank-4 viscosity tensor $\eta^{\alpha\beta\mu\nu}$, incorporate the anisotropy due to the magnetic field $B$, and then extract the shear viscosities $\eta_{xzxz}$ and $\eta_{xyxy}$ from Kubo's formula\footnote{For the sake of convenience, we will adopt the same conventions of those adopted in \cite{Huang:2011dc} and, thus, we will work in 4-dimensional Minkowski spacetime with mostly minus signature.}.
To clarify the discussion, we define the dissipation function $R$
\begin{equation}\label{eq:dissipation}
R = \frac{1}{2}\eta^{\mu\nu\alpha\beta}w_{\mu\nu}w_{\alpha\beta},
\end{equation}
where $w_{\mu\nu}=\frac{1}{2}\left(\nabla_\mu u_\nu+\nabla_\nu u_\mu \right)$, with $u^\mu$ being the 4-velocity and $\nabla_\mu= \Delta_{\mu\nu}\partial^\nu$; the object $\Delta_{\mu\nu}$ is just a projector on the directions orthogonal to $u^\mu$. Thus, the viscosity tensor gives us information about dissipation (i.e., generation of entropy) in the fluid. Taking the derivative of \eqref{eq:dissipation} with respect to $w_{\mu\nu}$, we obtain the usual stress tensor $\Pi^{\mu\nu}$
\begin{equation}
\Pi^{\mu\nu}= \eta^{\mu\nu\alpha\beta}w_{\alpha\beta}.
\label{onsagercondition}
\end{equation}

The construction of the viscosity tensor is based on its symmetry properties
\begin{equation}
 \eta^{\mu\nu\alpha\beta}(B)= \eta^{\nu\mu\alpha\beta}(B) =\eta^{\mu\nu\beta\alpha}(B)
\end{equation}
and the Onsager principle \cite{LandauKine,Huang:2011dc}
\begin{equation}
  \eta^{\mu\nu\alpha\beta}(B) =  \eta^{\alpha\beta\mu\nu}(-B).
\end{equation}
First, one writes down all the linear independent objects satisfying the above conditions of symmetry
\begin{align}
\text{(i)}& \ \Delta^{\mu\nu}\Delta^{\alpha\beta} \notag \\
\text{(ii)}& \ \Delta^{\mu\alpha}\Delta^{\nu\beta}+\Delta^{\mu\beta}\Delta^{\nu\alpha}  \notag \\
\text{(iii)}& \ \Delta^{\mu\nu}b^{\alpha}b^{\beta}+\Delta^{\alpha\beta}b^{\mu}b^{\nu}  \notag \\
\text{(iv)}& \ b^{\mu}b^{\nu}b^{\alpha}b^{\beta} \notag \\
\text{(v)} & \ \Delta^{\mu\alpha}b^{\nu}b^{\beta}+\Delta^{\mu\beta}b^{\nu}b^{\alpha}+\Delta^{\nu\alpha}b^{\mu}b^{\beta}+\Delta^{\nu\beta}b^{\mu}b^{\alpha}  \notag \\
\text{(vi)}& \ \Delta^{\mu\alpha}b^{\nu\beta}+\Delta^{\mu\beta}b^{\nu\alpha}+\Delta^{\nu\alpha}b^{\mu\beta}+\Delta^{\nu\beta}b^{\mu\alpha}  \notag \\
\text{(vii)}& \ b^{\mu\alpha}b^{\nu}b^{\beta}+b^{\mu\beta}b^{\nu}b^{\alpha}+b^{\nu\alpha}b^{\mu}b^{\beta}+b^{\nu\beta}b^{\mu}b^{\alpha}  
\end{align}
where $b^\mu$ is a spacelike vector orthogonal to the magnetic field, and $b^{\mu\nu}=\epsilon^{\mu\nu\alpha\beta}b^\alpha u^\beta$. This means that we have seven coefficients, five shear viscosities and two bulk viscosities. The shear viscosities are related to the traceless part of $\Pi^{\mu\nu}$ while the bulk viscosities are related to the trace of the stress tensor. We note that Onsager's condition in Eq.\ (\ref{onsagercondition}) is responsible for the presence of the two last tensors, (vi) and (vii), involving the Levi-Civita symbol $\epsilon^{\mu\nu\alpha\beta}$. These structures may appear in magnetized plasmas \cite{LandauKine,Huang:2011dc} but they are not present in the case of anisotropic superfluids \cite{LandauEla}.

In fact, according to \cite{Erdmenger:2010xm,Erdmenger:2012zu,Jain:2014vka}, for an anisotropic diagonal metric one can find only five linearly independent coefficients for the shear viscosity tensor due to metric fluctuations. This result is valid for the diagonal anisotropic background considered in this work, Eq. \eqref{eq:background}, and one can show using Kubo's formulas that the two coefficients associated with (vi) and (vii) trivially vanish due to the general structure of the background metric. 

For the sake of convenience, we will adopt the same combination of viscosity coefficients chosen in \cite{Huang:2011dc}. Thus, using the general linear combination of the structures above, we find the most general form of the viscosity tensor in the presence of a constant magnetic field
\begin{align}
\eta^{\mu\nu\alpha\beta} =& (-2/3\eta_0 +1/4\eta_1 +3/2\zeta_\perp)\text{(i)} + (\eta_0 )\text{(ii)} +(3/4\eta_1+3/2\zeta_\perp)\text{(iii)}\notag \\
  & +(9/4\eta_1 -4\eta_2 +3/2\zeta_\perp+3\zeta_\parallel )\text{(iv)} +(-\eta_2 )\text{(v)} +(-\eta_4)\text{(vi)} \notag \\
  & +(-\eta_3+\eta_4)\text{(vii)}.
\end{align}

The Kubo formulas for these coefficients are given by \cite{Huang:2011dc}
  \begin{align}\label{eq:kubo}
    \zeta_{\perp} &=-\frac{1}{3}\frac{\partial}{\partial \omega}\left[ 2G^{R}_{\tilde{P}_{\perp}\tilde{P}_{\perp}}(\omega,\vec{0})+G^{R}_{\tilde{P}_{\perp}\tilde{P}_{\parallel}}(\omega,\vec{0}) \right] \bigg\vert_{\omega\rightarrow 0} \notag \\
    \zeta_{\parallel} &= -\frac{1}{3}\frac{\partial}{\partial \omega}\left[ 2G^{R}_{\tilde{P}_{\perp}\tilde{P}_{\parallel}}(\omega,\vec{0})+G^{R}_{\tilde{P}_{\parallel}\tilde{P}_{\parallel}}(\omega,\vec{0}) \right] \bigg\vert_{\omega\rightarrow 0} \notag \\
    \eta_{0} &= -\frac{\partial}{\partial \omega}\text{Im}G^{R}_{\hat{T}^{12},\hat{T}^{12}}(\omega,\vec{0})\bigg\vert_{\omega\rightarrow 0} \notag \\
    \eta_{1} &= -\frac{4}{3}\eta_0 +2\frac{\partial}{\partial \omega}G^{R}_{\tilde{P}_{\perp}\tilde{P}_{\parallel}}(\omega,\vec{0}) \bigg\vert_{\omega\rightarrow 0} \notag\\
    \eta_{2} &= -\eta_{0} - \frac{\partial}{\partial \omega}\text{Im}G^{R}_{\hat{T}^{13},\hat{T}^{13}}(\omega,\vec{0})\bigg\vert_{\omega\rightarrow 0}\notag\\
    \eta_{3} &=- \frac{\partial}{\partial \omega}G^{R}_{\tilde{P}_{\perp},\hat{T}^{23}}(\omega,\vec{0})\bigg\vert_{\omega\rightarrow 0}  \notag\\
    \eta_{4} &=   - \frac{\partial}{\partial \omega}G^{R}_{\hat{T}^{13},\hat{T}^{23}}(\omega,\vec{0})\bigg\vert_{\omega\rightarrow 0},
  \end{align}
where $\tilde{P}_{\perp}= \hat{P}_{\perp}-(\theta_{\beta}-\Phi_\beta)\hat{\epsilon} $ and  $\tilde{P}_{\parallel}= \hat{P}_{\parallel}-\theta_{\beta}\hat{\epsilon} $; with  $\hat{P}_{\perp}=-\frac{1}{2}(\Delta_{\mu\nu}+ b_{\mu}b_{\nu})\hat{T}^{\mu\nu}$, $\hat{P}_{\parallel}=b_{\mu}b_{\nu}\hat{T}^{\mu\nu}$, $\theta_\beta=\left(\frac{\partial P}{\partial \epsilon} \right)_{B}$, $\Phi_{\beta}=-B\left(\frac{\partial M}{\partial \epsilon} \right)_{B}$, $\hat{\epsilon}=u_\mu u_\nu \hat{T}^{\mu\nu}$. $M$ is the magnetization of the plasma. Also, the retarded Green's function is defined as
\begin{equation}
G^{R}_{\hat{A} \hat{B}} (\omega, \vec{k}) = -i \int d^4x \, e^{-i k \cdot x} \theta(t) \left\langle \left[ \hat{A}(x), \hat{B}(0) \right] \right\rangle.
\end{equation}

The relation between these coefficients and the shear viscosities, $\eta_{\perp}$ and $\eta_{\parallel}$, calculated holographically, is
\begin{align}
\eta_0=\eta_\perp \notag \\
\eta_0+\eta_2= \eta_\parallel.
\end{align}

Finally, one can see from Eq. \eqref{eq:kubo} that for the type of background considered in this paper $\eta_3=\eta_4=0$ because the components $h_{xz}$, $h_{yz}$, $(h_{yy}+h_{xx})$, and $h_{zx}$ do not mix when one computes the action for the fluctuations. Thus, there are only five independents transport coefficients in this class of anisotropic backgrounds as mentioned above.


\end{document}